\documentclass[letter]{aa} 
\usepackage{graphicx}
\usepackage{natbib}
\bibpunct{(}{)}{,}{a}{}{;}
\usepackage{amssymb}
\usepackage{amsmath}
\usepackage{url}
\usepackage{units}
\usepackage{multirow,color}
\usepackage{txfonts}
\begin{document}

\title{Oxygen budget in low-mass protostars: the NGC1333-IRAS4A R1 shock observed in [\ion{O}{i}] at 63 $\mu$m with SOFIA-GREAT}
\titlerunning{Oxygen budget in low-mass protostars}


\author{L.E. Kristensen\inst{1}
\and A. Gusdorf\inst{2}
\and J.C. Mottram\inst{3}
\and A. Karska\inst{4}
\and R. Visser\inst{5}
\and H. Wiesemeyer\inst{6}
\and R. G{\"u}sten\inst{6}
\and R. Simon\inst{7}
}

\institute{
Centre for Star and Planet Formation, Niels Bohr Institute and Natural History Museum of Denmark, University of Copenhagen, {\O}ster Voldgade 5-7, DK-1350 Copenhagen K, Denmark, \email{lars.kristensen@nbi.ku.dk}  \and
LERMA, Observatoire de Paris, {\'E}cole normale sup{\'e}rieure, PSL Research University, CNRS, Sorbonne Universit{\'e}s, UPMC Univ. Paris 06, F-75231, Paris, France \and
Max Planck Institute for Astronomy, K{\"o}nigstuhl 17, 69117 Heidelberg, Germany \and
Centre for Astronomy, Nicolaus Copernicus University, Faculty of Physics, Astronomy and Informatics, Grudziadzka 5, PL-87100 Torun, Poland \and
European Southern Observatory, Karl-Schwarzschild-Strasse 2, 85748 Garching, Germany \and
Max-Planck-Institute for Radioastronomy, Auf dem H{\"u}gel 69, 53121 Bonn, Germany \and
I. Physikalisches Institut der Universit{\"a}t zu K{\"o}ln, Z{\"u}lpicher Strasse 77, 50937 K{\"o}ln, Germany
}

\date{Submitted: \today; Accepted: \today}


\abstract
{In molecular outflows from forming low-mass protostars, most oxygen is expected to be locked up in water. However, \textit{Herschel} observations have shown that typically an order of magnitude or more of the oxygen is still unaccounted for. To test if the oxygen is instead in atomic form, SOFIA-GREAT observed the R1 position of the bright molecular outflow from NGC1333-IRAS4A. The [\ion{O}{i}] 63 $\mu$m line is detected and spectrally resolved. From an intensity peak at +15 km\,s$^{-1}$ , the intensity decreases until +50 km\,s$^{-1}$. The profile is similar to that of high-velocity (HV) H$_2$O and CO 16--15, the latter observed simultaneously with [\ion{O}{i}]. A radiative transfer analysis suggests that $\sim$15\% of the oxygen is in atomic form toward this shock position. The CO abundance is inferred to be $\sim$10$^{-4}$ by a similar analysis, suggesting that this is the dominant oxygen carrier in the HV component. These results demonstrate that a large portion of the observed [\ion{O}{i}] emission is part of the outflow. Further observations are required to verify whether this is a general trend. 
}


\maketitle

\section{Introduction}

Oxygen is the third most abundant element in the Universe after hydrogen and helium. In the past few years, \textit{Herschel}-PACS observations have confirmed that the [\ion{O}{i}] 63$\mu$m line emission is one of the dominant gas cooling lines at far-infrared (FIR) wavelengths, along with CO and H$_2$O \citep[e.g.,][]{karska13, karska14, nisini15}. However, with a spectral resolution of $\sim$90~km\,s$^{-1}$, it is unclear where the [\ion{O}{i}] emission originates, even when considering velocity centroids \citep{dionatos16}. Only a handful of spectra show high-velocity line wings, indicating that only a small fraction of emission originates at $\varv$$>$100~km\,s$^{-1}$ \citep{nisini15}. The [\ion{O}{i}] line profile at lower velocities may be complex \citep[e.g.,][]{leurini15}, and it is clear that [\ion{O}{i}] emission does not follow more traditional outflow tracers such as low-$J$ CO emission \citep{mottram17}. 

Two other major reservoirs of oxygen exist in protostellar outflows: CO and H$_2$O \citep[OH is not believed to be a major reservoir,][]{wampfler13}. CO is an abundant and chemically stable molecule, and it is typically assumed that all volatile carbon is locked up in this molecule for a total abundance of $\sim$10$^{-4}$ \citep{yildiz13}, an assumption that holds for some protostellar outflows where both CO and H$_2$ are observed \citep{dionatos13}. The H$_2$O abundance, on the other hand, is more uncertain and has proved difficult to determine. Observations with \textit{Herschel}-HIFI suggest that the H$_2$O abundance is $\sim$ 10$^{-7}$--10$^{-5}$ toward outflow positions \citep[e.g.,][]{tafalla13, santangelo14, kristensen17}, which is orders of magnitude lower than $\sim$ 3$\times$10$^{-4}$, which is the expected value if all volatile oxygen not in CO is in H$_2$O \citep{vandishoeck14}. This low abundance of H$_2$O in outflows implies that our understanding of the total oxygen budget is incomplete. 

If the ``missing'' H$_2$O is in the form of atomic oxygen, then velocity-resolved spectra of H$_2$O and O should be similar, barring any large gradients in excitation conditions as a function of velocity. Furthermore, the total column density of H$_2$O, CO, and O should be equal to the total amount of volatile oxygen. To test these hypotheses, observations with the SOFIA-GREAT\footnote{GREAT is a development by the MPI f{\"u}r Radioastronomie and the KOSMA/Universit{\"a}t zu K{\"o}ln, in cooperation with the MPI f{\"u}r Sonnensystemforschung and the DLR Institut f{\"u}r Planetenforschung.} spectrometer were carried out toward the outflow spot NGC1333-IRAS4A R1, just north of the driving protostar. This region has been mapped with \textit{Herschel}-HIFI in several H$_2$O transitions \citep{santangelo14} and in [\ion{O}{i}] with \textit{Herschel}-PACS \citep{nisini15, dionatos16}. For these reasons, this position is an excellent initial target for determining the oxygen abundance and shedding light on where oxygen is in  outflows.

\section{Observations and results}
\label{sec:obs}

\begin{figure*}
\sidecaption
\includegraphics[width=12.5cm, angle=0]{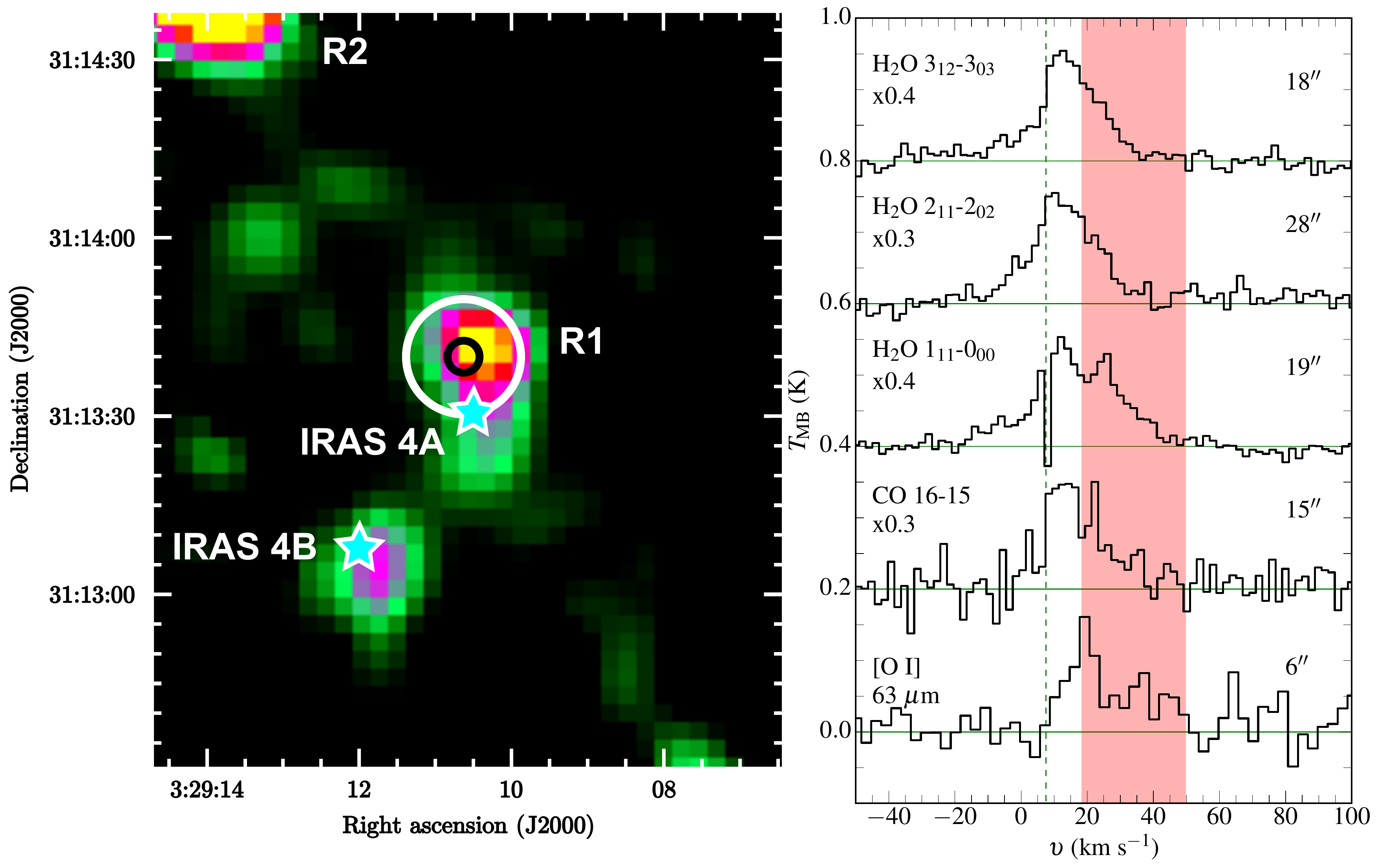}
\caption{\textit{Left:} [\ion{O}{i}] 63 $\mu$m integrated intensity map of the NGC1333-IRAS4A and 4B outflows as obtained with \textit{Herschel}-PACS \citep{nisini15}. The R1 position is shown with black and white circles just north of the source position; the smaller circle has a diameter of 6$''$, and the larger circle has a diameter of 20$''$. The protostellar positions are marked with stars. \textit{Right:} Spectra of selected transitions observed toward the R1 position in the NGC1333-IRAS4A outflow. The [\ion{O}{i}] and CO 16--15 spectra are observed with SOFIA-GREAT and the three H$_2$O spectra with \textit{Herschel}-HIFI. Most spectra have been scaled by a factor as labeled. The source velocity, 7.5 km\,s$^{-1}$ , is shown with a vertical dashed line, and the the HV component from 18--50 km\,s$^{-1}$ is highlighted in red. The beam size is shown on the right for each spectrum. 
\label{fig:spectrum}}
\end{figure*}

The Stratospheric Observatory for Infrared Astronomy \citep[SOFIA,][]{young12} with its German REceiver for Astronomy at Terahertz frequencies \citep[GREAT,][]{heyminck12} was used for the velocity-resolved observations of [\ion{O}{i}] presented in this study. They were carried out toward the bright [\ion{O}{i}] emission peak at the R1 position of the NGC1333-IRAS4A outflow \citep[RA: 3$^{\rm h}$29$^{\rm m}$10\fs63; Dec:$+$31$^\circ$13$'$40\farcs0; see Fig. \ref{fig:spectrum} and ][]{nisini15} on December 18, 2015. The total integration time was 26 min (on+off). The H-band receiver was tuned to the frequency of the [\ion{O}{i}] transition at 4744.777 GHz. The L2 receiver was tuned to the OH $^2\Pi_{1/2}$ $J$=3/2--1/2 triplet at 1834.760 GHz in the lower sideband. The intermediate frequency was set such that the CO $J$=16--15 transition at 1841.346 GHz appeared in the other sideband without blending with the OH line.

The raw data were calibrated to single-sideband antenna temperatures using the task \textsc{kalibrate} \citep{guan12}, which is a part of the \textsc{kosma} software. The spectra were then further reduced following standard procedures (including second-order baseline removal) using the \textsc{class} package\footnote{{\sc Class} is part of the {\sc Gildas} reduction package: \url{http://www.iram.fr/IRAMFR/GILDAS/}}. Individual scans were averaged using a 1/$\sigma_{\rm rms}$ weighting. Spectra were placed on the $T_{\rm MB}$ scale using a forward efficiency $\eta_{\rm f}$ of 0.97, and a main beam efficiency $\eta_{\rm MB}$ of 0.68 (L2) and 0.69 (H)\footnote{\url{http://www3.mpifr-bonn.mpg.de/div/submmtech/heterodyne/great/GREAT_calibration.html}}. The beam sizes at 4.7 and 1.8 THz are 6\farcs1 and 15\farcs3, respectively. 

The GREAT spectra are compared to HIFI spectral maps of several H$_2$O transitions from \citet{santangelo14}. These data were observed in 19--39$''$ beams, depending on the transition. For further reduction details, see \citet{santangelo14}. We note that the CO 16--15 spectrum obtained by \citet{santangelo14} was toward a different position, offset from the [\ion{O}{i}] peak by 3$''$W, 12$''$S, that is, offset by a full beam from the spectrum obtained with SOFIA-GREAT. For this reason, the higher $S/N$ CO 16--15 spectrum presented in \citet{santangelo14} is not used here. 

The reduced [\ion{O}{i}] and CO 16--15 spectra are shown in Fig. \ref{fig:spectrum}, together with the H$_2$O spectra  \citep{santangelo14}. OH was not detected and is not shown. Table \ref{tab:obs} contains the rms and intensities of the GREAT observations, including the OH limit. 

[\ion{O}{i}] emission traces a high-velocity (HV) component identified by \citet{santangelo14}, a component that peaks at $\varv_{\rm LSR}$$\sim$20--25 km\,s$^{-1}$. Although emission extends to lower velocities, there is no peak around the source velocity, as seen, for instance, in H$_2$O, and there is no narrow self-absorption as seen in the H$_2$O 1$_{11}$--0$_{00}$ transition (Fig. \ref{fig:spectrum}). This same component appears in CO 16--15, where the lower-velocity (LV) component is also identified, around $\varv_{\rm LSR}$$\sim$10--15 km\,s$^{-1}$. The HV component is most prominent in the H$_2$O 1$_{11}$--0$_{00}$ transition (Fig. \ref{fig:spectrum}) and not seen very clearly in the higher-$J$ 2$_{11}$--2$_{02}$ and 3$_{12}$--3$_{03}$ transitions. A small velocity shift of $\sim$5 km\,s$^{-1}$ may be seen between the HV component in H$_2$O, CO, and [\ion{O}{i}] profiles; however, given the low $S/N$ and large channel width (2--3 km\,s$^{-1}$), this shift is negligible. The H$_2$O spectra were obtained in large beams ($\sim$20$''$) and it is possible that emission from the source position may contaminate emission inherent to the R1 position. The on-source H$_2$O spectra do not show signs of an HV component \citep{kristensen12, mottram14}, suggesting that any contamination is negligible in the HV component.

To estimate the HV contribution to the CO 16--15 and H$_2$O transitions, the intensity is integrated between 18 and 50 km\,s$^{-1}$. The HV contribution is 30--50\% of the total integrated intensity for the H$_2$O transitions and 40\% for the CO 16--15 transition. All [\ion{O}{i}] emission is assigned to the HV component, as there is no component peaking at the source velocity. Finally, the GREAT spectra are compared to \textit{Herschel}-PACS fluxes \citep{karska13} extracted from a pixel in the same R1 location (see Appendix B for details, and Table \ref{tab:obs}). 

\begin{table}
\caption{Rms noise and total integrated intensity of the GREAT spectra.\label{tab:obs} }
\begin{center}
\footnotesize
\begin{tabular}{l c c c}\hline\hline 
Line & rms\tablefootmark{a} & $\int T_{\rm MB}$ d$\varv$ & PACS \\
& (mK) & (K\,km\,s$^{-1}$) & (K\,km\,s$^{-1}$) \\ \hline 
[\ion{O}{i}] 63$\mu$m & 33 & 2.7 & 1.0 \\
CO 16--15 & 77 & 11 & 9 \\
OH(1834 GHz) & 52 & $<$ 1.0\tablefootmark{b} & 1.2 \\
\hline\hline
\end{tabular}
\tablefoot{
\tablefoottext{a}{In 2 km\,s$^{-1}$ channels.}
\tablefoottext{b}{3$\sigma$ limit for a width of 20 km\,s$^{-1}$.}
}
\end{center}
\end{table}

\section{Analysis and discussion}
\label{sec:disc}

To constrain the oxygen budget, it is necessary to measure the column densities of the dominant oxygen-carriers, O, CO, and H$_2$O. H$_2$O has been observed in four transitions, so the strategy is to first derive excitation conditions for the HV component seen in H$_2$O, and next use these conditions to infer the O and CO column densities before discussing  possible implications. 

\subsection{Excitation conditions}

\begin{table}
\caption{Observed H$_2$O line ratios presented here and in other studies. \label{tab:ratio} }
\begin{center}
\footnotesize
\begin{tabular}{l c c c c c}\hline\hline 
Line ratio & Obs. & M14\tablefootmark{a} & B14\tablefootmark{b} & S12\tablefootmark{c} & S14\tablefootmark{d} \\ \hline 
1$_{10}$--1$_{01}$ / 1$_{11}$--0$_{00}$ & 0.73 & 0.75 & 1.32 & 1.18 & 0.79 \\
2$_{11}$--2$_{02}$ / 1$_{11}$--0$_{00}$ & 1.00 & 0.56 & 0.98 & 0.25 & 0.68 \\
3$_{12}$--3$_{03}$ / 1$_{11}$--0$_{00}$ & 0.60 & 0.50 & 0.16 & 0.13 & 0.26 \\
Least-squares &  & 0.20 & 0.54 & 0.99 & 0.22 \\
\hline\hline
\end{tabular}
\tablefoot{
\tablefoottext{a}{Average spot shock conditions, \citet{mottram14}.}
\tablefoottext{b}{L1157-B1, \citet{busquet14}.}
\tablefoottext{c}{L1448-R4 HV emission, \citet{santangelo12}.}
\tablefoottext{d}{NGC1333-IRAS4A R2, \citet{santangelo14}.}
}
\end{center}
\end{table}

\textit{Herschel}-HIFI observations of H$_2$O have been presented by a number of authors, both toward outflow spots and toward the source positions  where emission is still dominated by outflows \citep[e.g.,][]{santangelo12, santangelo14, busquet14, mottram14}. Because of the limited number of transitions available here and the uncertainty in separating emission from the HV component, the observed HV line ratios are first compared to those in other studies, and then the best-fit excitation conditions from those studies are used here. The best-fit excitation conditions are always inferred from simple non-local thermal equilibrium (non-LTE) models. 

Table \ref{tab:ratio} shows the line ratios measured in the HV component toward NGC1333-IRAS4A R1 compared to line ratios observed toward similar shock spots of low-mass protostellar outflows. A least-squares analysis shows that the line ratios observed here resemble those of NGC1333-IRAS4A R2 \citep{santangelo14} and the average ratios for spot shocks observed in a sample of 29 low-mass protostars \citep{mottram14}. \citeauthor{mottram14} find two solution regimes, one with a low density (3$\times$10$^5$ cm$^{-3}$), and one with a high density (10$^7$ cm$^{-3}$). The high-density solution is inconsistent with the [\ion{O}{i}] 63 $\mu$m / 145 $\mu$m ratio of 20, a ratio that suggests an H$_2$ density of 10$^5$ cm$^{-3}$ \citep{nisini15}. 

Both the H$_2$O and [\ion{O}{i}] line ratios are largely insensitive to temperature over the range of 300--1000 K, but CO 16--15 emission favors higher temperatures in the HV component \citep[e.g.,][]{lefloch12}. This leaves the H$_2$O column density and the size of the emitting region as the remaining unknowns. The excitation conditions inferred by \citet{santangelo14} and \citet{mottram14} probe different parameter regimes, and by comparing to these model results, we are effectively probing the full range of parameter space. The inferred H$_2$O excitation conditions from \citet{santangelo14} and \citet{mottram14} are summarized in Table \ref{tab:model}.

\subsection{Column densities and abundances}

\begin{table}
\caption{\textsc{Radex} model parameters and results for the R1 position. \label{tab:model} }
\begin{center}
\footnotesize
\begin{tabular}{l c c }\hline\hline 
Parameter & M14\tablefootmark{a} & S14\tablefootmark{b} \\ \hline 
$T$ (K) & 750 & 1000 \\
$n$(H$_2$) (cm$^{-3}$) & 3$\times$10$^5$ & 2$\times$10$^5$ \\
$r$ (AU)\tablefootmark{c} & 55 & 350 \\ 
$N$(H$_2$O) (cm$^{-2}$) & 1.0$\times$10$^{17}$ & 1.0$\times$10$^{16}$ \\ \hline
$N$(O) (cm$^{-2}$) & 1.5$\times$10$^{17}$ & 4.2$\times$10$^{16}$ \\
$N$(CO) (cm$^{-2}$) & 6.3$\times$10$^{17}$ & 1.7$\times$10$^{17}$ \\
$N$(OH) (cm$^{-2}$)\tablefootmark{d} & $<$5.0$\times$10$^{16}$ & $<$2.2$\times$10$^{16}$ \\
$X$(O)\tablefootmark{e} & 5.0$\times$10$^{-5}$ & 5.7$\times$10$^{-5}$ \\
$X$(H$_2$O)\tablefootmark{e} & 3.5$\times$10$^{-5}$ & 1.4$\times$10$^{-5}$ \\
$X$(CO)\tablefootmark{e} & 2.2$\times$10$^{-4}$ & 2.3$\times$10$^{-4}$ \\
C/O & 0.7 & 0.8 \\
\hline\hline
\end{tabular}
\tablefoot{Model input parameters are listed above the horizontal line and results below the line.
\tablefoottext{a}{Spot shock model results from \citet{mottram14}.}
\tablefoottext{b}{NGC1333-IRAS4A R2 model results from \citet{santangelo14}.}
\tablefoottext{c}{For a source distance of 235 pc \citep{hirota08}.}
\tablefoottext{d}{3$\sigma$ upper limit.}
\tablefoottext{e}{Abundances are calculated as $X$($Y$)/$X$(O$_{\rm tot}$) = $N$(Y)/[$N$(O)+$N$(CO)+$N$(H$_2$O)], where $X$(O$_{\rm tot}$)=3$\times$10$^{-4}$ is the abundance of volatile O w.r.t. H$_2$. Including the $N$(OH) upper limits will not significantly
change relative abundances. }
}
\end{center}
\end{table}

\begin{figure}
\center
\includegraphics[width=\columnwidth, angle=0]{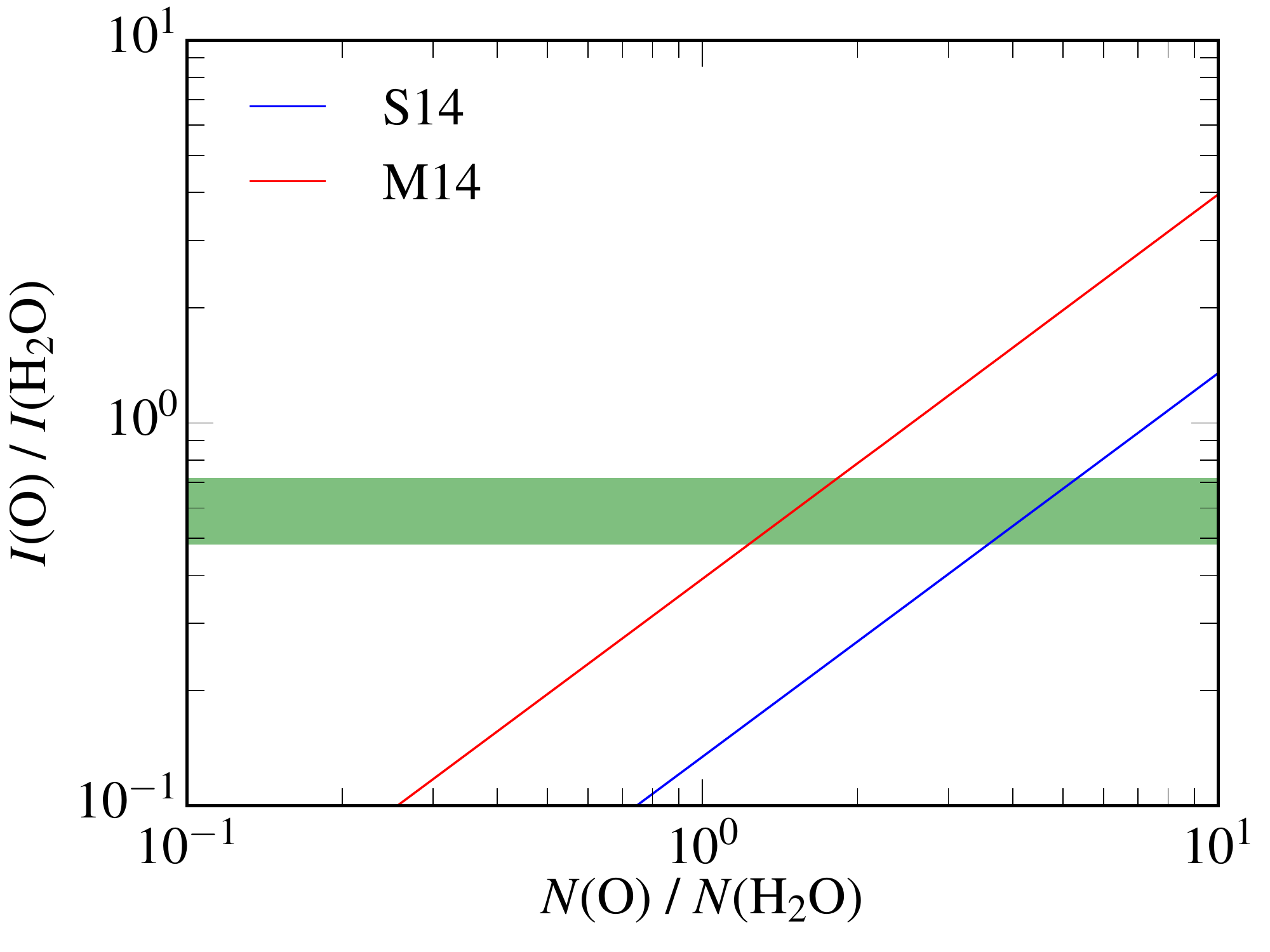}
\caption{\textsc{Radex} model predictions for the O/H$_2$O column density ratio inferred from the conditions in Table \ref{tab:model}, shown in red and blue, respectively. The observed line intensity ratio is plotted in green, and the ratio is against the H$_2$O 1$_{11}$--0$_{00}$ HV intensity of 4.7~K\,km\,s$^{-1}$.   
\label{fig:model}}
\end{figure}

Assuming that the excitation conditions inferred for the H$_2$O emission also apply to the HV emission in [\ion{O}{i}] and CO 16--15, their column densities may be inferred. To do so, the non-LTE radiative transfer code \textsc{radex} is used \citep{vandertak07}, with the collisional rate coefficients from \citet{daniel11} and \citet{wernli06}. The only free parameters in the modeling are the O and CO column densities, and the observational constraints are the CO 16--15 / H$_2$O 1$_{11}$--0$_{00}$ and [\ion{O}{i}] / H$_2$O 1$_{11}$--0$_{00}$ line ratios; the H$_2$O 1$_{11}$--0$_{00}$ line is chosen as the reference because it shows the HV component with the highest  signal-to-noise ratio $(S/N)$. Figure \ref{fig:model} shows the ratio of O/H$_2$O column density ratio vs. the observed line ratio; similar figures for the CO and OH column density ratios are shown in the Appendix, Fig. \ref{fig:model2}. The best-fit column densities are provided in Table \ref{tab:model}. A similar exercise may be performed for the OH upper limit using the collisional rate coefficients from \citet{offer94}, with the results also shown in Fig. \ref{fig:model}. 

To translate the inferred column densities into abundances, a total abundance of volatile oxygen of 3$\times$10$^{-4}$ is assumed, and the abundance is calculated as $X$($Y$)/$X$(O$_{\rm tot}$) = $N$(Y)/[$N$(O)+$N$(CO)+$N$(H$_2$O)], where $X$(O$_{\rm tot}$)=3$\times$10$^{-4}$ is the abundance of volatile O with respect to H$_2$ \citep{sofia04}. This results in an abundance of atomic oxygen of 5--6$\times$10$^{-5}$, or about 25\% of that of CO (see Table \ref{tab:model}). These calculations do not include the upper limits on OH, but if these were included, the resulting abundances would change by $<$ 10\%, or in other words, not significantly. The upper limit on OH translates to $X$(OH) $<$ 10$^{-5}$, and so this species is most likely not a significant carrier of oxygen.

\subsection{Implications}

These results demonstrate that the abundance of atomic oxygen is significantly lower than the total abundance of volatile oxygen of $\sim$ 3$\times$10$^{-4}$ \citep{whittet10} in the HV component. The total fraction of atomic oxygen is $\sim$15\% in this component, depending on the assumed excitation conditions. Such a low abundance suggests that the bulk of the gas is molecular, and any calculation of shock and mass-loss properties based on the assumption that the oxygen abundance is 3$\times$10$^{-4}$ is thus a lower limit by up to a factor of six \citep{nisini15}. We note that the [\ion{O}{i}] spectrum does not show any signs of additional components compared to what is seen in H$_2$O line profiles \citep{santangelo14}, particularly, there are no signs of extremely high velocity (EHV) components seen at high angular resolution in SiO emission closer to the source \citep{santangelo15}. This conclusion currently only applies to a single off-source shock spot. Further observations will tell if this conclusion holds toward other shocks as well.

The CO abundance is $\sim$2$\times$10$^{-4}$, which is entirely consistent with most volatile carbon locked up in this molecule \citep[cf. $X$(CO)=3$\times$10$^{-4}$ w.r.t. H$_2$;][]{whittet10}. Moreover, this abundance is consistent with combined observations of CO and H$_2$ toward multiple shocked regions in Serpens, where the CO abundance with respect to H$_2$ was measured to be 0.5--2.5$\times$10$^{-4}$ \citep{dionatos13}, based on a direct comparison of high-$J$ CO and rotational H$_2$ emission. Thus, CO is most likely a good proxy for the total mass in the high-temperature part of the HV component, as long as the correct transition is used. 

The H$_2$O abundance is notoriously difficult to measure, primarily because the denominator of the $N$(H$_2$O)/$N$(H$_2$) ratio is difficult to measure directly \citep{vandishoeck14, kristensen17}. One method, as outlined here, is to circumvent H$_2$ and instead account for the dominant oxygen carriers in the outflow, and then compare these to the total amount of volatile oxygen available. This analysis confirms previous abundance estimates using CO 16--15 as a proxy for the H$_2$ column density, finding a total H$_2$O abundance of a few times 10$^{-5}$ \citep{santangelo14, kristensen17}. 

The inferred results depend strongly on the assumed excitation conditions, particularly whether the excitation conditions inferred from the H$_2$O excitation analysis apply to CO 16--15 and [\ion{O}{i}], and by implication if the excitation conditions used here are optimal. If the different species occupy different spatial regions of the outflow shock, for example, the same excitation conditions need not apply. The line profiles of particularly CO 16--15 and [\ion{O}{i}] appear so similar in the HV component that it seems likely that these two lines trace the same region. The HV component is partly folded into the H$_2$O line profiles, and extracting the HV component from the high-$J$ H$_2$O line profiles leaves room for interpretation. Furthermore, at least part of the underlying HV emission seen in H$_2$O may originate from the source position, where broad HV line wings are also
seen very clearly \citep{kristensen12, mottram14}. These uncertainties propagate into the inferred excitation conditions, but by using ``standard'' H$_2$O excitation conditions, some of this uncertainty is alleviated. 

\section{Summary and conclusions}

Weak [\ion{O}{i}] 63 $\mu$m emission is detected and velocity-resolved toward the R1 position of the NGC1333-IRAS4A outflow. The line profile is remarkably similar to the HV components of H$_2$O and CO 16--15, suggesting that all three species trace the same gas. When we used a radiative-transfer analysis, 15\% of the oxygen was found to be in atomic form toward the HV component at this outflow shock position, meaning that the bulk of material is in molecular form. The dominant molecular carrier of oxygen is CO. The H$_2$O abundance is confirmed to be low, a few times 10$^{-5}$, similar to O. [\ion{O}{i}] is not detected in the LV component of this shock position, suggesting that the bulk of material in this component is molecular, consistent with the excitation results presented by \citet{santangelo14}. 

This single [\ion{O}{i}] spectrum already sheds light on the oxygen budget toward low-mass outflows. Most oxygen at high velocities is in molecular form, primarily CO and to a lesser degree H$_2$O. Clearly, the next steps include making a survey of protostellar outflow positions, and also observing [\ion{O}{i}] toward the source positions, where a similarly low H$_2$O abundance has been inferred \citep{mottram14, kristensen17}.

\begin{acknowledgements}
The authors would like to thank S. Cabrit for stimulating discussions, G. Sandell for help with scheduling the observations, and the referee for valuable comments that helped improving the clarity of the paper. Submillimeter astronomy in Copenhagen is supported by the European Research Council (ERC) under the European Union's Horizon 2020 research and innovation programme (grant agreement No 646908) through ERC Consolidator Grant ``S4F''. Research at the Centre for Star and Planet Formation is funded by the Danish National Research Foundation. JCM acknowledges support from the European Research Council under the European Community's Horizon 2020 framework program (2014-2020) via the ERC Consolidator grant `From Cloud to Star Formation (CSF)' (project number 648505). AK acknowledges support from the Polish National Science Center grants 2013/11/N/ST9/00400 and 2016/21/D/ST9/01098. Based on observations made with the NASA/DLR Stratospheric Observatory for Infrared Astronomy (SOFIA). SOFIA is jointly operated by the Universities Space Research Association, Inc. (USRA), under NASA contract NAS2-97001, and the Deutsches SOFIA Institut (DSI) under DLR contract 50 OK 0901 to the University of Stuttgart.
\end{acknowledgements}

\bibliographystyle{aa}
\bibliography{30310_ap}

\appendix

\section{Line data and observational details}

\begin{table*}[!t]
\caption{Transition properties. \label{tab:mol} }
\begin{center}
\footnotesize
\begin{tabular}{l l c c c l c}\hline\hline 
Species & Transition & Freq. & $E_{\rm up}/k_{\rm B}$ & $A$\tablefootmark{a} & Instr. & Beam \\
&& (GHz) & (K) & (s$^{-1}$) & & ($''$) \\ \hline
O & $^3$P$_1$--$^3$P$_2$ & 4744.777 & 228 & 8.91(--5) & GREAT & 6.1 \\
CO & 16--15 & 1841.346 & 750 & 4.05(--4) & GREAT & 15.3 \\
OH & $\Omega,J,P$ = $\frac{1}{2}$--$\frac{1}{2}$, $\frac{3}{2}$--$\frac{1}{2}$, -- -- + & 1834.760 & 270 & 6.45(--2) & GREAT & 15.3 \\
H$_2$O & 1$_{10}$--1$_{01}$ & 556.936 & 61 & 3.46(--3) & HIFI & 38.1 \\
& 1$_{11}$--0$_{00}$ & 1113.343 & 53 & 1.84(--2) & HIFI & 19.0 \\
& 2$_{11}$--2$_{02}$ & 752.033 & 137 & 7.06(--3) & HIFI & 29.2 \\
& 3$_{12}$--3$_{03}$ & 1097.365 & 249 & 1.65(--2) & HIFI & 19.3 \\
\hline\hline
\end{tabular}
\tablefoot{
\tablefoottext{a}{$A(B)$ = $A$ $\times$ 10$^{B}$.}
}
\end{center}
\end{table*}

Table \ref{tab:mol} presents the properties of the transitions observed or presented here. 

\section{Complementary PACS data}

\textit{Herschel}-PACS fluxes are extracted from the R1 position following the method of \citet{karska13}, where data reduction details may also be found. The footprint obtained as part of the `Water in Star-forming regions with \textit{Herschel}' \citep[WISH,][]{vandishoeck11} key program overlaps with the R1 position to better than 1$''$, and no regridding of the spatially undersampled data is necessary. The extracted PACS fluxes are provided in Table \ref{tab:pacs}, which also includes the [\ion{O}{i}] 145$\mu$m line flux. 

The PACS resolution below wavelengths of $\sim$ 150 $\mu$m is limited by the pixel size and not the diffration limit\footnote{Figure 8 in the PACS Spectroscopy Performance and Calibration document, \url{http://herschel.esac.esa.int/twiki/pub/Public/PacsCalibrationWeb/PacsSpectroscopyPerformanceAndCalibration_v3_0.pdf}}. A flux of 2.36$\times$10$^{-16}$ W\,m$^{-2}$ thus corresponds to 1.0 K\,km\,s$^{-1}$. To compare this value to the observed GREAT flux, we assume that the emission is coming from a point-like region and scale the GREAT flux by the relative difference in beam- or aperture size. The beam dilution of 2.7 K\,km\,s$^{-1}$ as observed with GREAT (Table 1) in a 6\farcs1 beam to a 9\farcs4 $\times$ 9\farcs4 aperture is 2.7 K\,km\,s$^{-1}$ 1.13 $\times$ 6\farcs1$^2$ / 9\farcs4$^2$ = 1.3 K\,km\,s$^{-1}$, where the factor of 1.13 accounts for the Gaussian beam size of GREAT. The difference between 1.0 and 1.3 K km/s is negligible in light of calibration uncertainties, which are typically on the order of 10--15\%.

\begin{table}[!t]
\caption{\textit{Herschel}-PACS fluxes of selected species and transitions toward the R1 position \citep{karska13}. \label{tab:pacs} }
\begin{center}
\footnotesize
\begin{tabular}{l c}\hline\hline 
Line & Flux (10$^{-16}$ W m$^{-2}$) \\ \hline 
[\ion{O}{i}] 63$\mu$m & 2.36$\pm$0.09 \\
\phantom{}[\ion{O}{i}] 145$\mu$m & 0.11$\pm$0.01 \\
CO 16--15 & 1.43$\pm$0.02 \\
OH(1838 GHz) & 0.10$\pm$0.02 \\
OH(1834 GHz) & 0.20$\pm$0.02 \\
\hline\hline
\end{tabular}
\end{center}
\end{table}

\section{CO and OH excitation conditions}

\begin{figure}
\center
\includegraphics[width=\columnwidth, angle=0]{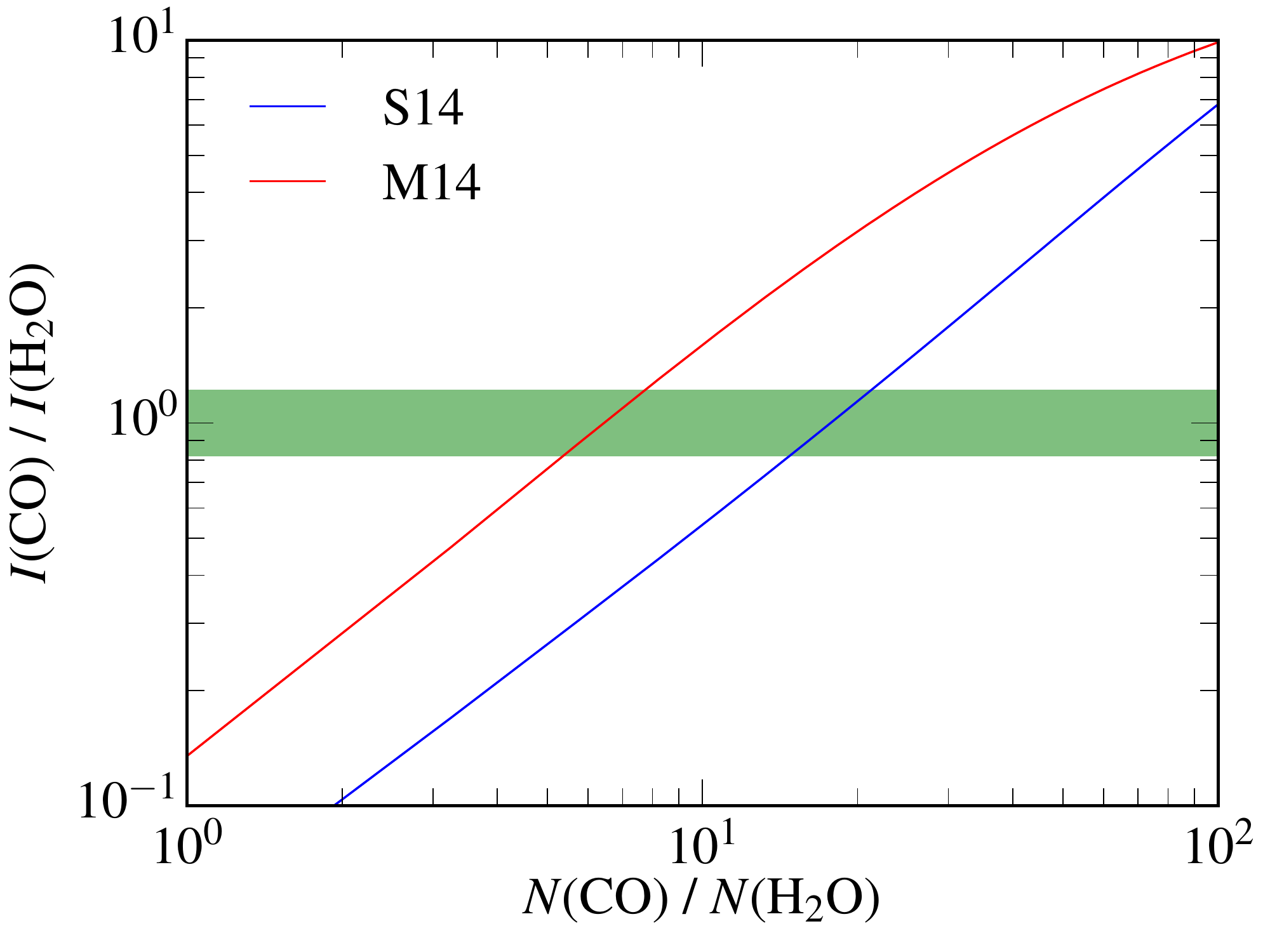}
\includegraphics[width=\columnwidth, angle=0]{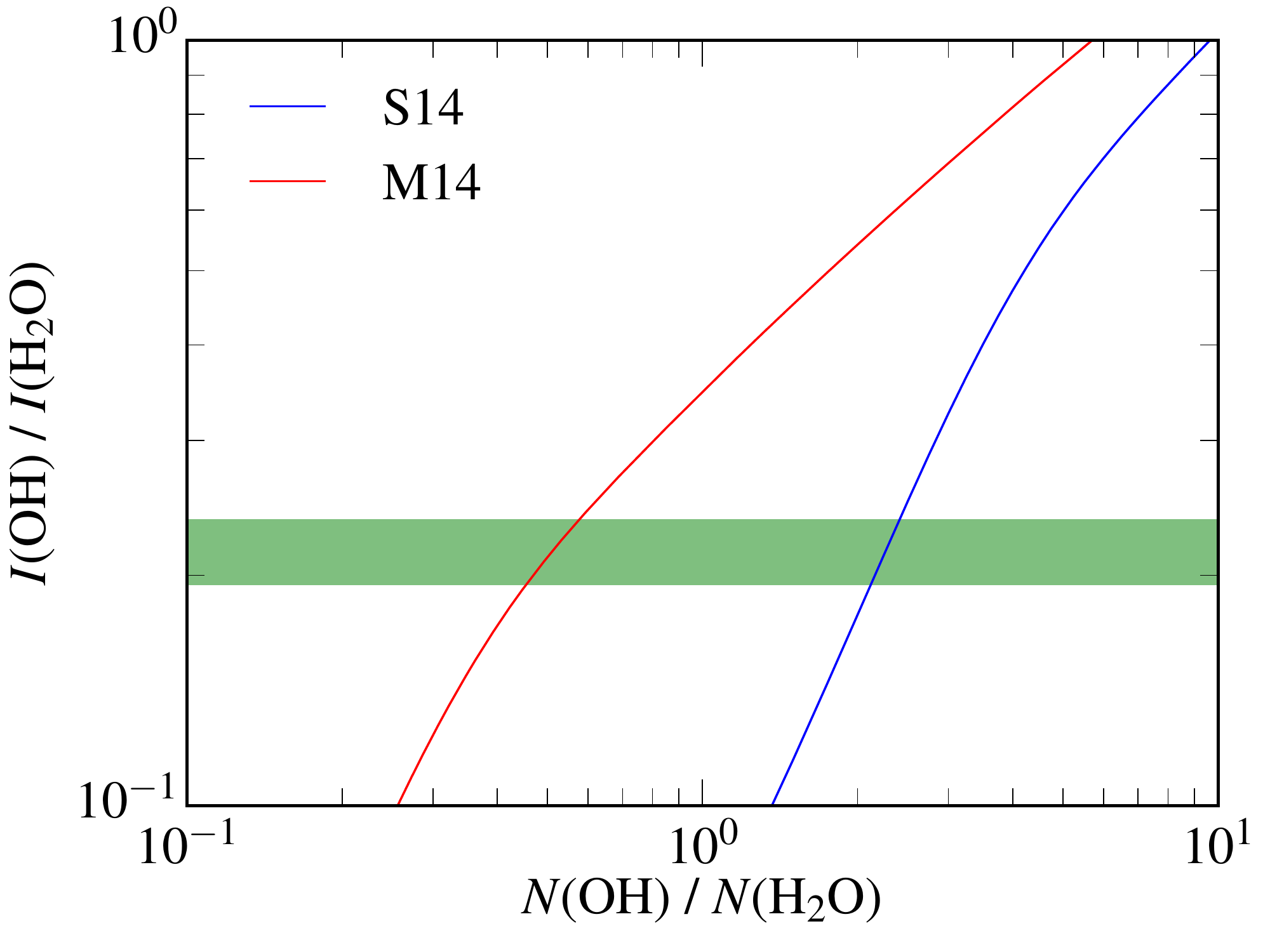}
\caption{\textsc{Radex} model predictions for the CO/H$_2$O (\textit{top}), and OH/H$_2$O (\textit{bottom}) column density ratios inferred from the conditions in Table \ref{tab:model}, shown in red and blue, respectively. The observed line intensity ratios are plotted in green, where the OH/H$_2$O ratio is an upper limit, and all ratios are against the H$_2$O 1$_{11}$--0$_{00}$ HV intensity of 4.7~K\,km\,s$^{-1}$.   
\label{fig:model2}}
\end{figure}

Figure \ref{fig:model2} shows the ratio of the CO and OH  column densities over that of H$_2$O vs. the observed line ratio. The best-fit column densities are provided in Table \ref{tab:model}.

\end{document}